\documentclass[journal,12pt,onecolumn,draftclsnofoot,]{IEEEtran}
\usepackage{algorithm}
\usepackage{algorithmicx}
\usepackage{algpseudocode}
\usepackage{amsmath}
\usepackage{amssymb}
\usepackage{array}
\usepackage{arydshln}
\usepackage{bm}
\usepackage{booktabs}
\usepackage{braket}
\usepackage{cite}
\usepackage{enumitem}
\usepackage{graphicx}
\usepackage[mathscr]{euscript}
\usepackage{mathtools}
\usepackage{multirow}
\usepackage{framed} 
\usepackage[framed]{ntheorem}
\usepackage{nicefrac}
\usepackage{pgfplots}
\usepackage{qtree}
\usepackage{adjustbox}
\usepackage{rotating}
\pgfplotsset{compat = 1.13}
\usepackage{stfloats}
\usepackage{subfig}
\usepackage{url}
\usepackage{tikz}
\usepackage{tkz-berge}
\usepackage[framemethod=TikZ]{mdframed}
\usetikzlibrary{shapes, backgrounds,calc, snakes}
\usetikzlibrary{decorations.pathreplacing}
\usepackage{fancybox}

\tikzstyle{vertex} = [circle, draw, inner sep = 0pt, minimum size = 10pt]
\newcommand{\vertex}{\node[vertex]}
\newframedtheorem{frm-prop}{Property}

\definecolor{bblue}{rgb}{0.12392, 0.0490, 0.9588}
\definecolor{sskyblue}{rgb}{0.1529, 0.5882, 0.9216}
\definecolor{ggreen}{rgb}{0.5020, 0.7961, 0.3451}
\definecolor{yyellow}{rgb}{0.9765, 0.9804, 0.0784}

\definecolor{color0}{HTML}{FF0147}
\definecolor{color1}{HTML}{F400DC}
\definecolor{color2}{HTML}{BA0DFF}
\definecolor{color3}{HTML}{5700E8}
\definecolor{color4}{HTML}{0B03FF}
\definecolor{color5}{HTML}{0957F4}
\definecolor{color6}{HTML}{03B3FF}
\definecolor{color7}{HTML}{08E8DA}
\definecolor{color8}{HTML}{07FF8E}
\definecolor{color9}{HTML}{51FF0A}

\IEEEoverridecommandlockouts \IEEEpubid{\makebox[\columnwidth]{978-1-5386-3531-5/17/\$31.00 ©2017 European Union \hfill} \hspace{\columnsep}\makebox[\columnwidth]{ }}
\begin{document}

\title{Graph-Based Resource Allocation with Conflict Avoidance for V2V Broadcast Communications}

\author{\IEEEauthorblockN{Luis F. Abanto-Leon}
\IEEEauthorblockA{Department of Electrical Engineering\\
Eindhoven University of Technology\\
Email: l.f.abanto@tue.nl}
\and
\IEEEauthorblockN{Arie Koppelaar}
\IEEEauthorblockA{NXP Semiconductors\\
	Eindhoven\\
	Email: arie.koppelaar@nxp.com}
\and
\IEEEauthorblockN{Sonia Heemstra de Groot}
\IEEEauthorblockA{Department of Electrical Engineering\\
	Eindhoven University of Technology\\
	Email: sheemstradegroot@tue.nl}}

\maketitle

\begin{abstract}
	In this paper we present a graph-based resource allocation scheme for sidelink broadcast vehicle-to-vehicle (V2V) communications. Harnessing available information on the geographical position of vehicles and spectrum resources utilization, eNodeBs are capable of allotting the same set of sidelink resources to several different vehicles in order for them to broadcast their signals. Hence, vehicles sharing the same resources would ideally be in different communications clusters for the interference level---generated due to resource repurposing---to be maintained under control. Within a communications cluster, it is crucial that vehicles transmit in orthogonal time resources to prevent conflicts as vehicles---with half-duplex radio interfaces---cannot transmit and receive simultaneously. In this research, we have envisaged a solution based on a bipartite graph, where vehicles and spectrum resources are represented by vertices whereas the edges represent the achievable rate in each resource based on the signal--to--interference--plus--noise ratio (SINR) that vehicles perceive. The aforementioned constraint on time orthogonality of allocated resources can be approached by aggregating conflicting vertices into macro-vertices which, in addition, narrows the search space yielding a solution with computational complexity equivalent to the conventional graph matching problem. We show mathematically and through simulations that the proposed approach yields an optimal solution. In addition, we provide simulations showing that the proposed method outperforms other competing approaches, specially in scenarios with high vehicular density.  
\end{abstract}

\IEEEpeerreviewmaketitle

\section{Introduction}
Vehicle--to--vehicle (V2V) communications is one of the novel use cases under the umbrella of the next generation of wireless systems 5G. How can V2V communications---in its many facets---be leveraged to comply with the very stringent latency and reliability requirements that this type of scenario poses, has attracted much interest. In standardization groups, for instance, support of time-critical communications in safety-related applications has drawn superlative attention due to the immediate implications. In this context, several studies have led to the conclusion that connectivity-enabled vehicles have the potential to prevent accidents \cite{b1}.

In V2V $\textit{Mode 3}$, vehicles are assigned sidelink resources---by an eNodeB---to periodically broadcast their signals, namely cooperative awareness messages (CAMs) \cite{b2}. CAM messages contain important information about the vehicle, e.g., velocity, direction, position, which can be availed by other vehicles (or drivers) for better decision-making. An important aspect in the resource allocation process is to guarantee that vehicles---within the same communications cluster---will broadcast their signals in orthogonal time resources. This is due to the fact that their half-duplex interface does not allow simultaneous transmission and reception. As a result, vehicles in the same cluster must be allocated resources in different subframes to prevent conflicts \cite{b7}. Nevertheless, a resource serving a vehicle in certain communications cluster can be repurposed by another provided that the latter vehicle is exclusively associated to a different cluster. Thus, eNodeBs will play an important role in $(i)$ effectively allocating resources to vehicles in coverage and $(ii)$ inferring knowledge about the association of vehicles to in-coverage clusters. It is worth clarifying that although resource allocation is managed by eNodeBs, the fact that vehicles can communicate directly without data having to traverse eNodeBs is beneficial due to proximity gain \cite{b4}, lower latency and resource reuse gain.

On the other hand, matching is a fundamental problem in combinatorial optimization and has found applications in a plethora of areas. Several matching problems can be represented by graphs, which may exhibit a wide variety of morphologies and different degrees of connectedness. A very specific type of matching problems can be modeled as weighted bipartite graphs, where the objective is to find a vertex--to--vertex matching---between elements of two disjoint sets of vertices---that attains a maximum sum of edge weights. This classical problem is called herein \textit{unconstrained weighted graph matching} \cite{b9} \cite{b10}. As already mentioned, resource allocation for V2V communications has a primary time orthogonality requirement to prevent conflicts. A solution based on a bipartite graph would provide sound framework for approaching a problem of this kind. However, due to additional conflict constraints, the \textit{unconstrained weighted graph matching} approach cannot be applied straightforwardly to our problem. We have envisaged a solution where the graph with additional conflict constraints---called herein \textit{constrained weighted graph matching}---is transformed into a simpler problem that can fundamentally be approached as an unconstrained graph.

The objective of this paper is two-fold: $(i)$ prove that an optimal solution for the \textit{constrained weighted graph matching} problem exists by means of two different approaches and $(ii)$ discuss the suitability of such an approach for avoiding resource allocation conflicts in broadcast vehicular communications. Our paper is organized in the following manner. In section II we explain the motivation for our work and synthesize our contributions. In Section III, we briefly revisit the \textit{unconstrained weighted graph matching} problem. In Section IV, our proposed approach is described in detail. Section V is devoted for discussing simulation results. Finally, in Section VI, we summarize our conclusions.
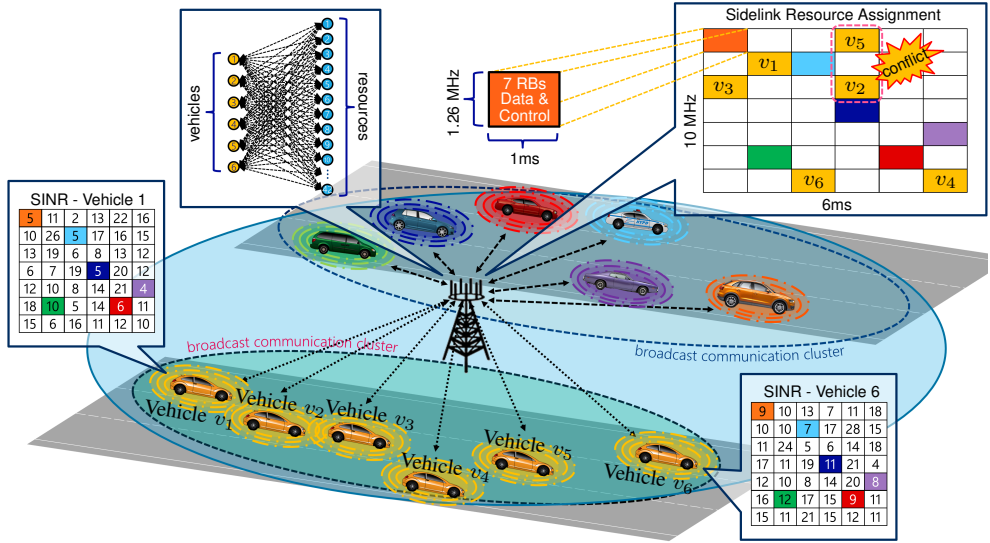
\begin{figure*}[!t]
	\begin{center}
		\begin{tikzpicture}
			\node (img) {\includegraphics[width=0.8\linewidth]{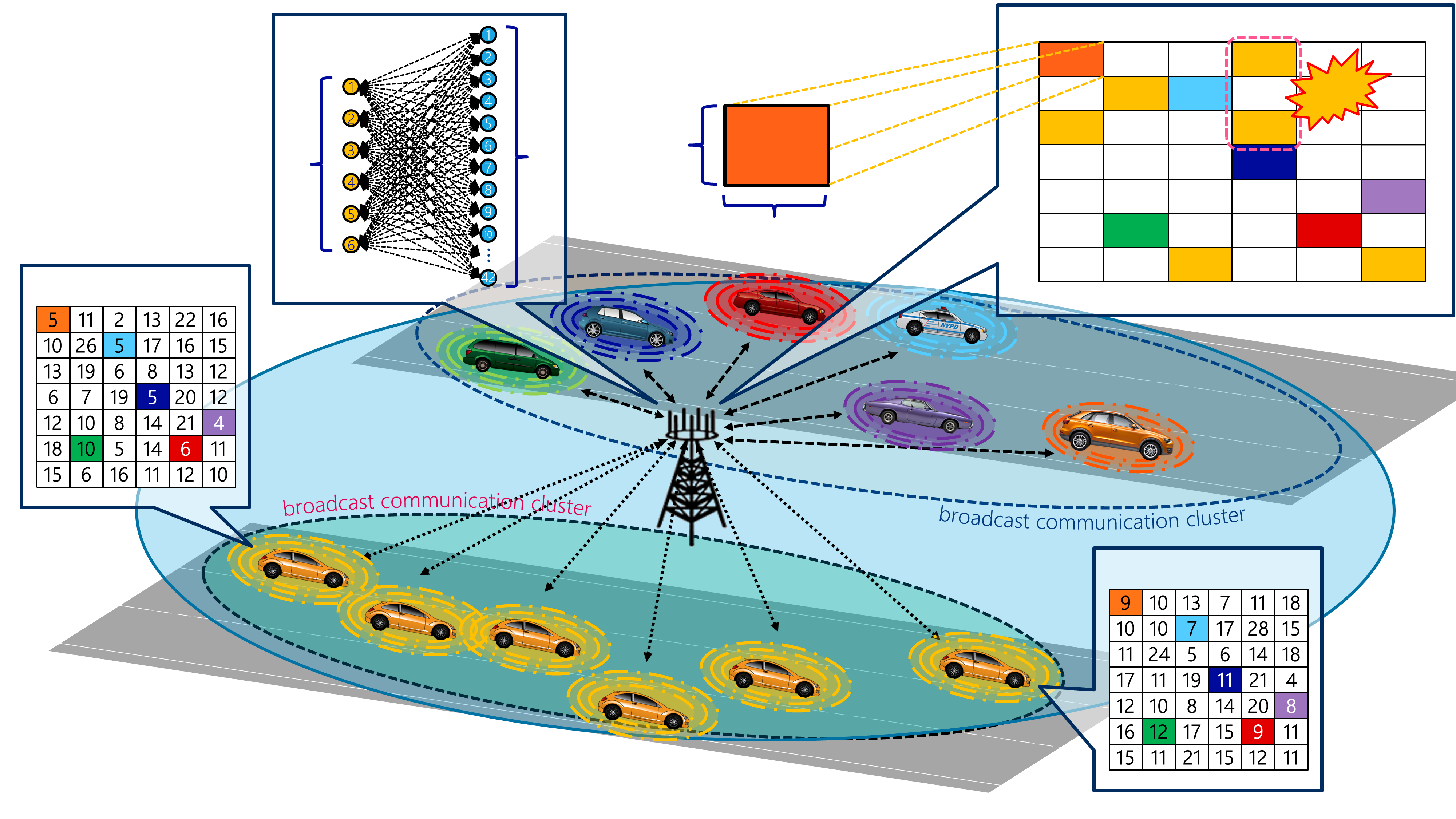}};
			\node at (-5.35, 1.07) {\tiny \textsf{SINR - Vehicle 1 }};
			\node at (4.4, -1.49) {\tiny \textsf{SINR - Vehicle 6 }};
			
			\node[rotate= -9] at (-4.0, -1.8) {\scriptsize Vehicle $v_1$};
			\node[rotate= -9] at (-2.8, -1.65) {\scriptsize Vehicle $v_2$};
			\node[rotate= -9] at (-1.6, -1.8) {\scriptsize Vehicle $v_3$};
			\node[rotate= -9] at (-0.6, -2.5) {\scriptsize Vehicle $v_4$};
			\node[rotate= -9] at (0.5, -2.2) {\scriptsize Vehicle $v_5$};
			\node[rotate= -9] at (2.1, -2.65) {\scriptsize Vehicle $v_6$};
			
			\node at (3.1, 2.55) {\scriptsize $v_3$};
			\node at (3.7, 2.85) {\scriptsize $v_1$};
			\node at (4.3, 1.3) {\scriptsize $v_6$};
			\node at (4.85, 3.15) {\scriptsize $v_5$};
			\node at (4.85, 2.55) {\scriptsize $v_2$};
			\node at (6.05, 1.3) {\scriptsize $v_4$};
			
			\node[rotate= 90] at (-3.95, 2.2) {\tiny \textsf{vehicles}};
			\node[rotate= -90] at (-1.7, 2.3) {\tiny \textsf{resources}};
			\node[rotate= 25] at (5.5, 2.88) {\tiny \textsf{conflict}};
			
			\node[color = white] at (0.45, 2.6) {\tiny \textsf{7 RBs}};
			\node[color = white] at (0.45, 2.4) {\tiny \textsf{Data \&}};
			\node[color = white] at (0.45, 2.18) {\tiny \textsf{Control}};
			
			\node at (0.45, 1.6) {\tiny \textsf{1ms}};
			\node[rotate = 90] at (-0.5, 2.35) {\tiny \textsf{1.26 MHz}};
			
			\node at (4.6, 1) {\tiny \textsf{6ms}};
			\node[rotate = 90] at (2.65, 2.1) {\tiny \textsf{10 MHz}};
			
			\node at (4.55, 3.5) {\tiny \textsf{Sidelink Resource Assignment}};
		\end{tikzpicture}
		\caption{Vehicular broadcast communications scenario in mode-3 via sidelink}
		\label{fig1}
	\end{center}
	\vspace{-0.5cm}
\end{figure*}

\section{Motivation and Contributions}
In our system model, we consider that sidelink spectrum resources for V2V communications are decoupled from uplink/downlink bands. Thus, the scenario herein is different from the underlay configuration \cite{b5} where idle uplink resources are opportunistically utilized for sidelink vehicular communications. The reason is that for safety-related applications, it will be necessary to count with a frequency band that is always accessible and does not depend on the availability of shared resources. 

As we already mentioned, the allocated resources must be orthogonal in time domain, otherwise conflicts will arise. Our motivation is to develop an approach capable of dealing with such type of constraints in order to provide dependable communications. For instance, in Fig. \ref{fig1}. we observe two clusters, each consisting of 6 vehicles. In one of the clusters, we can observe a resource conflict where vehicles $v_2$ and $v_5$ have been allotted resources in the same time subframe. Our goal is to harness the information harvested by vehicles---such as channel conditions--- for the eNodeB to perform an efficient and effective resource allocation task.

The contributions of our work are summarized:
\begin{itemize}
	\item Kuhn-Munkres \cite{b6} is a computationally efficient method that can be used for solving resource allocation problems formulated as bipartite graphs. However, due to additional time orthogonality constraints, the resultant problem is not directly approachable by the aforementioned method. In our solution, vertices conflicting among each other have been aggregated into macro-vertices yielding a resultant graph which is solvable by Kuhn-Munkres.
	\item Vertex aggregation virtually cuts down the number of effective vertices and therefore narrows the number of potential solutions without affecting optimality. The envisaged approach can attain an optimal solution at the same computational expense as the \textit{unconstrained weighted graph matching} problem.
	\item We show through simulations that our approach is capable of providing fairness among all vehicles, especially in scenarios with high vehicle density.
\end{itemize}

\section{Unconstrained Weighted Graph Matching}
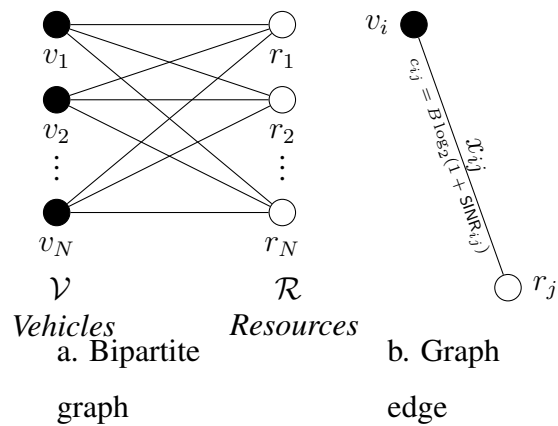
\begin{figure}[!b]
	\centering
	\[\begin{tikzpicture}
	\vertex[fill] (v1) at (0,-0.5) [label = below:$v_{1}$] {};
	\vertex[fill] (v2) at (0,-1.5) [label = below:$v_{2}$] {};
	\vertex[fill] (v3) at (0,-3) [label = below:$v_{N}$] {};
	\vertex (r1) at (3,-0.5) [label = below:$r_{1}$] {};
	\vertex (r2) at (3,-1.5) [label = below:$r_{2}$] {};
	\vertex (r3) at (3,-3) [label = below:$r_{N}$] {};
	
	\path
	(v1) edge (r1)
	(v1) edge (r2)
	(v1) edge (r3)
	(v2) edge (r1)
	(v2) edge (r2)
	(v2) edge (r3)
	(v3) edge (r1)
	(v3) edge (r2)
	(v3) edge (r3);
	
	\node[text width = 0.2cm] at (0,-4) {$\mathcal{V}$};
	\node[text width = 0.2cm] at (-0.5,-4.5) {\textit{Vehicles}};
	\node[text width = 0.2cm] at (3,-4) {$\mathcal{R}$};
	\node[text width = 0.2cm] at (2.4,-4.5) {\textit{Resources}};
	\node at (0,-2.3) {\vdots};	
	\node at (3,-2.3) {\vdots};
	
	\node[text width = 3cm] at (1.5,-5.25) {a. Bipartite graph};
	
	\vertex[fill] (vi) at (4.75,-0.5) [label = left:$v_{i}$] {};
	\vertex (rj) at (6,-4) [label = right:$r_{j}$] {};
	
	\path
	(vi) edge (rj);
	
	\node[rotate = -70.26] at (5.2,-2.25) {\tiny $c_{ij} = B \log_2(1 + \mathsf{SINR}_{ij})$};
	\node[rotate = -70.26] at (5.55,-2.25) {$x_{ij}$};
	
	\node[text width = 2.2cm] at (5.5,-5.25) {b. Graph edge};
	
	\end{tikzpicture}\]
	\caption{Unconstrained Weighted Bipartite Graph}
	\label{fig2}
\end{figure}

A weighted complete bipartite graph $G = (\mathcal{V}, \mathcal{R}, \mathcal{E})$ consists of two disjoint sets of vertices $\mathcal{V}$, $\mathcal{R}$ and a set $\mathcal{E} = \mathcal{V} \times \mathcal{R}$ of edges, as depicted in Fig. \ref{fig2}. An edge $x_{ij}$ connects a vertex $v_i \in \mathcal{V}$ with a vertex $r_j \in \mathcal{R}$ and has an associated weight $c_{ij}$. The objective is to find a matching $\mathcal{M} \subseteq \mathcal{E}$ that associates every vertex in $\mathcal{V}$ with a vertex in $\mathcal{R}$---in a one--to--one manner---and attaining maximum sum of weights. In a bipartite graph, when the cardinality of the vertex sets are equal, i.e.,  $\vert \mathcal{R} \vert = \vert \mathcal{V} \vert = N$, a perfect matching can be attained.

In Fig. \ref{fig2}, the vertices $v_i$ represent the vehicles that belong to the same communications cluster $\mathcal{V}$ whereas the vertices $r_j$ represent the allotable resources, which are denoted by $\mathcal{R}$. In this paper, we consider that the edge weights $c_{ij}$ represent the achievable rate on each resource based on the SINR that vehicles perceive, i.e., $c_{ij} = B \log_2(1 + \mathsf{SINR}_{ij})$, where $B$ is the bandwidth of a sidelink resource. The goal is to assign each vehicle $v_i$---in the several communications clusters that may exist---a resource $r_j$ for it to broadcast its signal with the aim of maximizing the sum-rate capacity of the system.


\subsection{Summation Representation}
The bipartite graph matching problem is expressed by
\begin{subequations} \label{e1}
	\begin{gather} 
	\begin{align}
		& {\rm max}~{\displaystyle \sum_{i = 1}^N \sum_{j = 1}^N c_{ij} x_{ij}} \\
		\vspace{-0.2cm} 
		& {\rm subject~to} \nonumber \\
			& {\displaystyle \sum_{i = 1}^N x_{ij} = 1}, \hspace{0.2cm} {j = 1, 2, \dots, N} \\
			\vspace{-0.25cm} 
			& {\displaystyle \sum_{j = 1}^N x_{ij} = 1}, \hspace{0.2cm} {i = 1, 2, \dots, N} \\
			\vspace{-0.25cm} 
			& x_{ij} = \{0, 1\}, \hspace{0.2cm} {i,j = 1, 2, \dots, N} 
	\end{align}
	\end{gather}
\end{subequations}
where constraint (1b) guarantees that each vertex $v_i \in \mathcal{V}$ will be matched to one vertex $r_j \in \mathcal{R}$ only. The constraint (1c) ensures that each vertex $r_j \in \mathcal{R}$ will be associated with a single vertex $v_i \in \mathcal{V}$. Thus, both constraints enforce a one--to--one matching. The constraint (1c) ensures that $x_{ij}$ is either 1---if vertex $v_i \in \mathcal{V}$ is matched to vertex $r_j \in \mathcal{R}$---or 0 if they are unmatched. An optimal solution can be effectively found by means of Kuhn-Munkres algorithm \cite{b6} with $\mathcal{O}(\max({\vert \mathcal{V} \vert, \vert \mathcal{R} \vert})^3) = \mathcal{O}(N^3)$ complexity.

\subsection{Matrix Representation}
An alternative representation of (1) is given by (\ref{e2})
\begin{subequations} \label{e2}
	\begin{gather} 
	\begin{align}
	\hspace{-2cm}
		& {\rm max} ~ {\bf c}^T {\bf x} & {\bf c} \in \mathbb{R}^{M},  {\bf x} \in \mathbb{B}^M,\\
		& {\rm subject~to}~ {\bf Ax} = {\bf 1}  & {\bf A} \in \mathbb{B}^{2N \times M} 
	\end{align}
\end{gather}
\end{subequations}
where $M = N^2$, $\mathbb{R}$ denotes the real numbers and $\mathbb{B}$ represents the $\{0, 1\}$ realm. The totally unimodular matrix $\mathbf{A}$ encapsulates $2N$ constraints---$N$ constraints due to vertices in $\mathcal{V}$ and $N$ additional constraints that arise due to vertices in $\mathcal{R}$. Finally, $\mathbf{x} = [x_{1,1}, \dots, x_{N,N}]^T$, $\mathbf{c} = [c_{1,1}, \dots, c_{N,N}]^T$ are the solution vector and weight vector, respectively.

\section{Proposed Constrained Weighted Graph Matching Solution}
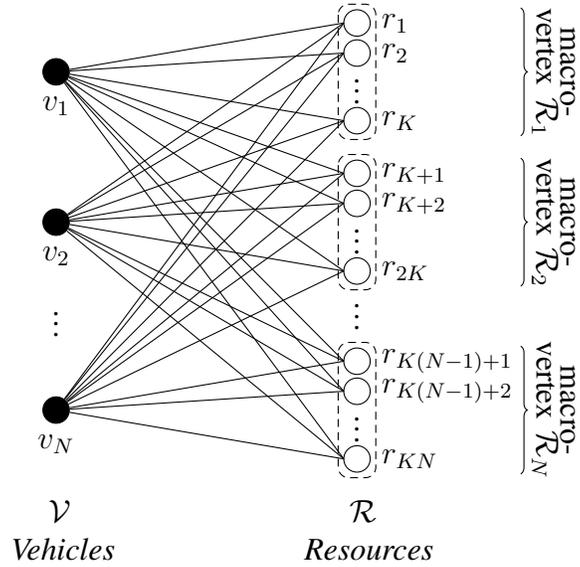
\begin{figure}[!t]
	\centering
	\[\begin{tikzpicture}
	
	\vertex[fill] (v1) at (0,-1.25) [label = below:$v_{1}$] {};
	\vertex[fill] (v2) at (0,-3.25) [label = below:$v_{2}$] {};
	\node at (0,-4.5) {\vdots};
	\vertex[fill] (v3) at (0,-5.75) [label = below:$v_{N}$] {};
	
	\vertex (r11) at (4,-0.6) [label = right:$r_{1}$] {};
	\vertex (r12) at (4,-1) [label = right:$r_{2}$] {};
	\node at (4,-1.4) {\vdots};
	\vertex (r13) at (4,-1.9) [label = right:$r_{K}$] {};
	
	\vertex (r21) at (4,-2.6) [label = right:$r_{K+1}$] {};
	\vertex (r22) at (4,-3) [label = right:$r_{K+2}$] {};
	\node at (4,-3.4) {\vdots};
	\vertex (r23) at (4,-3.9) [label = right:$r_{2K}$] {};
	
	\node at (4,-4.4) {\vdots};
	
	\vertex (r31) at (4,-5.1) [label = right:$r_{K(N-1)+1}$] {};
	\vertex (r32) at (4,-5.5) [label =right:$r_{K(N-1)+2}$] {};
	\node at (4,-5.9) {\vdots};
	\vertex (r33) at (4,-6.4) [label = right:$r_{KN}$] {};
	
	\path
	(v1) edge (3.825,-0.6)
	(v1) edge (3.825,-1)
	(v1) edge (3.825,-1.9)
	(v1) edge (3.825,-2.6)
	(v1) edge (3.825,-3)
	(v1) edge (3.825,-3.9)
	(v1) edge (3.825,-5.1)
	(v1) edge (3.825,-5.5)
	(v1) edge (3.825,-6.4)
	
	(v2) edge (3.825,-0.6)
	(v2) edge (3.825,-1)
	(v2) edge (3.825,-1.9)
	(v2) edge (3.825,-2.6)
	(v2) edge (3.825,-3)
	(v2) edge (3.825,-3.9)
	(v2) edge (3.825,-5.1)
	(v2) edge (3.825,-5.5)
	(v2) edge (3.825,-6.4)
	
	(v3) edge (3.825,-0.6)
	(v3) edge (3.825,-1)
	(v3) edge (3.825,-1.9)
	(v3) edge (3.825,-2.6)
	(v3) edge (3.825,-3)
	(v3) edge (3.825,-3.9)
	(v3) edge (3.825,-5.1)
	(v3) edge (3.825,-5.5)
	(v3) edge (3.825,-6.4);
	
	\draw[densely dashed,rounded corners=4]($(r11)+(-.25,.25)$)rectangle($(r13)+(0.25,-.25)$);
	\draw[densely dashed,rounded corners=4]($(r21)+(-.25,.25)$)rectangle($(r23)+(0.25,-.25)$);
	\draw[densely dashed,rounded corners=4]($(r31)+(-.25,.25)$)rectangle($(r33)+(0.25,-.25)$);
	
	\draw[decoration={brace, raise=5pt},decorate] (6,-0.4) -- node[right=6pt] {} (6,-2.1);
	\draw[decoration={brace, raise=5pt},decorate] (6,-2.4) -- node[right=6pt] {} (6,-4.1);
	\draw[decoration={brace, raise=5pt},decorate] (6,-4.9) -- node[right=6pt] {} (6,-6.6);
	
	\node[rotate=-90] at (6.8,-1.25) {macro-};
	\node[rotate=-90] at (6.8,-3.25) {macro-};
	\node[rotate=-90] at (6.8,-5.75) {macro-};
	\node[rotate=-90] at (6.5,-1.25) {vertex $\mathcal{R}_1$};
	\node[rotate=-90] at (6.5,-3.25) {vertex $\mathcal{R}_2$};
	\node[rotate=-90] at (6.5,-5.75) {vertex $\mathcal{R}_N$};
	
	\node[text width = 0.2cm] at (0,-7.1) {$\mathcal{V}$};
	\node[text width = 0.2cm] at (-0.5,-7.6) {\textit{Vehicles}};
	\node[text width = 0.2cm] at (4,-7.1) {$\mathcal{R}$};
	\node[text width = 0.2cm] at (3.4,-7.6) {\textit{Resources}};
	
	\end{tikzpicture}\]
	\caption{Constrained Weighted Bipartite Graph}
	\label{fig3}
	\vspace{-0.5cm}
\end{figure}

Let $G = (\mathcal{V}, \mathcal{R}, \mathcal{E})$ be a bipartite graph such that the cardinality of the sets $\mathcal{V}$ and $\mathcal{R}$ are related by $\vert\mathcal{R}\vert = K \vert\mathcal{V}\vert = KN$, as depicted in Fig. \ref{fig3}. In this scheme, the $KN$ vertices in $\mathcal{R}$ are grouped into $N$ disjoint groups $\{\mathcal{R}_{\alpha}\}_{\alpha = 1}^N$ called macro-vertices, such that $\mathcal{R} = \cup_{\alpha = 1}^{N} \mathcal{R}_{\alpha}$, $\mathcal{R}_{\alpha} \cap \mathcal{R}_{\alpha'} = \emptyset$, $\forall \alpha \neq \alpha'$. Each macro-vertex $\mathcal{R}_{\alpha}$ is an aggregation of $K$ vertices, i.e., $\vert\mathcal{R}_{\alpha}\vert = K$. The target is to find a vertex--to--vertex matching with maximum sum of weights such that no two vertices in $\mathcal{V}$ are matched to any two vertices that belong to the same macro-vertex $\mathcal{R}_{\alpha}$. This condition must be satisfied as it portrays the time orthogonality requirement that prevents allocation conflicts. Notice that this type of constraints for conflict avoidance cannot be inherently managed by the approach described in Section III. Thus, this motivated us for developing an approach capable of handling such constraints. We will show via two different approaches in Section IV.A and Section IV.B, that the optimal solution is tantamount to finding the maximum vertex--to--macro-vertex matching. 

The problem with additional macro-vertex constraints is formulated in (\ref{e3})
\begin{subequations} \label{e3}
	\begin{gather} 
	\begin{align}
			& {\rm max}~{\displaystyle \sum_{i = 1}^N \sum_{j = 1}^{KN} c_{ij} x_{ij}} \\
			\vspace{-0.2cm} 
			& {\rm subject~to} \nonumber \\
				& {\displaystyle \sum_{j = 1}^{KN} x_{ij} = 1}, \hspace{0.2cm} {i = 1, 2, \dots, N} \\
				\vspace{-0.25cm}
				& {\displaystyle \sum_{i = 1}^N \sum_{j \in \mathcal{R}_{\alpha}} x_{ij} = 1}, \hspace{0.2cm} {\alpha = 1, 2, \dots, N} \\
				\vspace{-0.25cm}
				& x_{ij} = \{0, 1\}, \hspace{0.2cm} {i,j = 1, 2, \dots, N} 
	\end{align}
\end{gather}
\end{subequations}
where the constraint (3b) enforces every vertex $v_i \in \mathcal{V}$ to be matched to a single vertex $r_j \in \mathcal{R}$. The constraint (3c) stipulates that in any macro-vertex $\mathcal{R}_{\alpha}$, only one of its constituting vertices $r_j \in \mathcal{R}_{\alpha} $ can be matched to a single vertex $v_i \in \mathcal{V}$. 

In Fig. \ref{fig3}, $K$ represents the number of resources per subframe, each with duration 1 ms \cite{b3}. Thus, a macro-vertex $\mathcal{R}_{\alpha}$ represents the aggregation of all the resources in subframe $\alpha$. $N$ denotes the number of available subframes in which the resource allocation task can be accomplished. For example, in Fig. \ref{fig1}, the number of resources per subframe is $K = 7$ whereas the amount of subframes is $N = 6$, which yields a total of 42 resources. Since we consider that the system does not operate at overload, the number of subframes in the system should be at least equal to the cardinality of the maximum-cardinality cluster. Otherwise, there will exist vehicles that will not be served. When the number of vehicles in a cluster is smaller than the number of subframes, dummy vehicles can be added such that $\vert\mathcal{R}\vert = K \vert\mathcal{V}\vert = KN$.

\subsection{Solution Derivation using Summation Representation}
We will show that the matching problem with macro-vertex constraints can be recast as an unconstrained graph with smaller cardinality and thus can be solved at the same computational complexity expense as the \textit{unconstrained weighted graph matching} problem.

We observe that for any two edges $x_{ij}$ and $x_{ik}$ that share the same vertex $v_i$, the following holds true: $x_{ij} x_{ik} = 0, ~ $ if $ ~ j \neq k ~ $and$ ~ r_j, r_k \in \mathcal{R}_{\alpha}, \forall \alpha $. The validity of this expression can be readily verified because every vertex $v_i \in \mathcal{V}$ can be matched to one vertex $r_j \in \mathcal{R}$ only. Thus, harnessing the previous relation, the following expression also holds true: $\sum_{j \in \mathcal{R}_{\alpha}} \sum_{\substack {k \in \mathcal{R}_{\alpha} \\ j \neq k}} x_{ij}x_{ik} = 0$ for any $v_i \in \mathcal{V}$, $r_j, r_k \in \mathcal{R}_{\alpha}, \forall \alpha$. Furthermore, a generalized result is given by $\sum_{i = 1}^N \sum_{\alpha = 1}^N \sum_{j \in \mathcal{R}_{\alpha}} \sum_{\substack {k \in \mathcal{R}_{\alpha} \\ j \neq k}} c_{ij}x_{ij}x_{ik} = 0$. Notice that adding $c_{ij}$ to the zero-product $x_{ij}x_{ik}$ does not affect the result. On the other hand, the cost function in (3a) can be expressed as $\sum_{i = 1}^N \sum_{j = 1}^{KN} c_{ij} x_{ij} = \sum_{i = 1}^N \sum_{\alpha = 1}^{N} \sum_{j \in \mathcal{R}_{\alpha}} c_{ij} x_{ij}$. Moreover, since $x_{ij} = \{0,1\}~\forall i, j$, the cost function can be further simplified employing the quadratic terms as $\sum_{i = 1}^N \sum_{\alpha = 1}^{N} \sum_{j \in \mathcal{R}_{\alpha}} c_{ij} x_{ij} = \sum_{i = 1}^N \sum_{\alpha = 1}^{N} \sum_{j \in \mathcal{R}_{\alpha}} c_{ij} x_{ij}^2$. In addition, notice that constraint (3a) can be equivalently expressed as $\sum_{j = 1}^{KN} x_{ij} = \sum_{\alpha = 1}^{N} \sum_{j \in \mathcal{R}_{\alpha}} x_{ij} = 1$. Thus, collecting the previous outcomes, the problem in (\ref{e3}) can be recast as (\ref{e4})
\begin{subequations} \label{e4}
	\begin{gather} 
	\small
	\begin{align}
		& {\rm max}~{\displaystyle \sum_{i = 1}^N \sum_{\alpha = 1}^N \sum_{j \in \mathcal{R}_{\alpha}} c_{ij} x_{ij}^2 + \sum_{i = 1}^N \sum_{\alpha = 1}^N \sum_{j \in \mathcal{R}_{\alpha}} \sum_{\substack {k \in \mathcal{R}_{\alpha} \\ j \neq k}} c_{ij}x_{ij}x_{ik}} \\
		\vspace{-0.5cm} 
		& {\rm subject~to} \nonumber \\
			& {\displaystyle \sum_{\alpha = 1}^{N} \sum_{j \in \mathcal{R}_{\alpha}} x_{ij} = 1}, \hspace{0.2cm} {i = 1, 2, \dots, N} \\
			\vspace{-0.25cm}
			& {\displaystyle \sum_{i = 1}^N \sum_{j \in \mathcal{R}_{\alpha}} x_{ij} = 1}, \hspace{0.2cm} {\alpha = 1, 2, \dots, N} \\
			\vspace{-0.25cm}
			& x_{ij} = \{0, 1\}, \hspace{0.2cm} {i,j = 1, 2, \dots, N} 
	\end{align}
\end{gather}
\end{subequations}

It was feasible to augment the cost function by adding zero-valued terms since the solution optimality would not be affected. Thus, the new cost function is
\begin{equation} \label{e5}
	\small
	\begin{array}{l}
	
		~{\displaystyle \sum_{i = 1}^N \sum_{\alpha = 1}^N \sum_{j \in \mathcal{R}_{\alpha}} c_{ij} x_{ij}^2 + \sum_{i = 1}^N \sum_{\alpha = 1}^N \sum_{j \in \mathcal{R}_{\alpha}} \sum_{\substack {k \in \mathcal{R}_{\alpha} \\ j \neq k}} c_{ij}x_{ij}x_{ik}} \\

		\vspace{-0.2cm} \\
		
		= {\displaystyle \sum_{i = 1}^N \sum_{\alpha = 1}^{N} \underbrace{\Bigg(\sum_{j \in \mathcal{R}_{\alpha}} c_{ij} x_{ij}\Bigg)}_\text{$d_{i \alpha}$} \underbrace{\Bigg(\sum_{k \in \mathcal{R}_{\alpha}} x_{ik}\Bigg)}_\text{$y_{i \alpha}$}} \\
		
		\vspace{-0.2cm} \\
	\end{array}
\end{equation}

Employing the definition of $y_{i \alpha}$, the constraint (4b) can be expressed as ${\sum_{\alpha = 1}^{N} \sum_{j \in \mathcal{R}_{\alpha}} x_{ij} = \sum_{\alpha = 1}^{N} y_{i \alpha} = 1}$ whereas the constraint (4c) is reduced to ${\sum_{i = 1}^N \sum_{j \in \mathcal{R}_{\alpha}} x_{ij} = \sum_{i = 1}^{N} y_{i \alpha} = 1}$.

Finally, stemming from the fact that $y_{i \alpha}$ is the outcome of a sum on binary variables $x_{ij}$, it yields that $y_{i \alpha}$ is also binary. Upon collecting the previous results, the original problem in (\ref{e3}) can be expressed as (\ref{e6})
\begin{subequations} \label{e6}
	\begin{gather} 
	\begin{align}
		& {\rm max}~{\displaystyle \sum_{i = 1}^N \sum_{\alpha = 1}^N d_{i \alpha} y_{i \alpha}} \\
		\vspace{-0.2cm}
		& {\rm subject~to} \nonumber \\
			& {\displaystyle \sum_{i = 1}^N y_{i \alpha} = 1}, \hspace{0.2cm} {\alpha = 1, 2, \dots, N} \\
			\vspace{-0.25cm}
			& {\displaystyle \sum_{\alpha = 1}^N y_{i \alpha} = 1}, \hspace{0.2cm} {i = 1, 2, \dots, N} \\
			\vspace{-0.25cm}
			& y_{i \alpha} = \{0, 1\}, \hspace{0.2cm} {i, \alpha = 1, 2, \dots, N} 
	\end{align}
\end{gather}
\end{subequations}

Note that (\ref{e6}) makes evident that a matching will be between vertices $v_i \in \mathcal{V}$ and macro-vertices $\mathcal{R}_{\alpha}$, which is equivalent to solving the unconstrained matching problem in (\ref{e1}). However, $d_{i \alpha} = \sum_{j \in \mathcal{R}_{\alpha}} c_{ij} x_{ij}$ are unknown as they depend on $x_{ij}$. ITo remove the dependency of $d_{i \alpha}$ on $x_{ij}$, and be able to solve (\ref{e6}), in the following we explain our reasoning. We will show that we can obtain a solution $y_{i \alpha}$ for (\ref{e6}) without first solving for $x_{ij}$.

\underline{\bf Analysis:} \vspace{0.15cm}

If there exists an optimal solution $\{x_{i j}^{\star}\}_{i = 1, j = 1}^{i = N, j = N}$ to (\ref{e3}), then the following holds true
\begin{itemize}
	\item There must also exist a solution $\{y_{i \alpha}^{\star}\}_{i = 1,\alpha = 1}^{i = N,\alpha = N}$ that satisfies (\ref{6}) optimally. This is because, $x_{ij}$ can be linearly mapped to $y_{i \alpha} = \sum_{j \in \mathcal{R}_{\alpha}} x_{ij}$.
	\item Thus, given an optimal solution $\{y_{i \alpha}^{\star}\}_{i=1,\alpha=1}^{i=N,\alpha=N}$, for some vertex $v_{i'}$ and macro-vertex $\mathcal{R}_{\alpha'}$, the following is provable.
	\begin{enumerate}
		\item If some $y_{i' \alpha'}^{\star} = 0$, this implies that the edge $y_{i' \alpha'}^{\star}$ is unmatched with vertex $v_{i'}\in \mathcal{V}$ with macro-vertex $\mathcal{R}_{\alpha'}$. This is equivalent to asserting that vertex $v_{i'} \in \mathcal{V}$ is not matched to any of the vertices $r_j\in \mathcal{R}_{\alpha'}$.
		\item If some $y_{i' \alpha'}^{\star} = 1$, there must exist an edge $x_{i'j'} = 1$ that matches vertex $v_{i'} \in \mathcal{V}$ with vertex $r_{j'} \in \mathcal{R}_{\alpha'}$. As a consequence, such said edge must be optimal, i.e., $x_{i'j'}^{\star} = x_{i'j'} = 1$, because $y_{i' \alpha'}^{\star} =  x_{i'j'}^{\star} + \sum_{\substack{j \in \mathcal{R}_{\alpha'} \\ j \neq j'}} x_{i'j}^{\star} = 1$.
	\end{enumerate}
	\item If some $x_{i'j'}^{\star} = 1$, then its associated weight $c_{i'j'}$ is also involved in the optimal solution. In other words, $d_{i' \alpha'} = c_{i'j'} x_{i'j'}^{\star} + \sum_{\substack{j \in \mathcal{R}_{\alpha'} \\ j \neq j'}} c_{i'j} x_{i'j}^{\star} = c_{i'j'}$. To wit, $d_{i' \alpha'}$ will be either $c_{i'j'}, \exists r_{j'} \in \mathcal{R}_{\alpha'}$ when $x_{i'j'}^{\star} = 1$ or 0 when $x_{i'j'}^{\star} = 0$. However, note that if a maximum matching exists, it will then be attained regardless of the other values $c_{ij}$ as long as every $c_{i'j'} \geq c_{ij}, \forall i \neq i', j \neq j'$. Thus, without loss of optimality, $d_{i' \alpha'} = \max\{c_{i'j} \lvert j \in \mathcal{R}_{\alpha'}\}$.
\end{itemize}
\hfill $\blacksquare$

We have shown that it is possible to remove the dependency of $d_{ij}$ on $x_{ij}$ and thus (\ref{e6}) can be solved as an \textit{unconstrained weighted graph matching} problem.

\subsection{Solution Derivation using Matrix Representation}
In this section, we develop a generalized framework by which it is possible to show that (\ref{e2}) is a particular case of the vertex aggregation case, i.e., the \textit{unconstrained weighted graph matching} problem when $K = 1$. Thus, the problem is formulated as
\begin{subequations} \label{e7}
	\vspace{-0.05cm}
	\begin{gather} 
	\begin{align}
	& {\rm max} ~ {\bf c}^T {\bf x} \\ 
	& {\rm subject~to}~
	{\left[
		\begin{array}{c}
		{\bf I}_{N \times N} \otimes {\bf 1}_{1 \times N}\\
		\hline
		{\bf 1}_{1 \times N} \otimes {\bf I}_{N \times N} 
		\end{array}
		\right]} \otimes {\bf 1}_{1 \times K} ~ {\bf x} = {\bf 1} 
	\end{align}
	\end{gather}
\end{subequations}
where $\otimes$ represents the tensor product operator, ${\bf c} \in \mathbb{R}^{M},  {\bf x} \in \mathbb{B}^M$ with $M = KN^2$. 

Because the solution $\bf x$ exists on the binary realm, the cost function (7a) can be equivalently expressed as ${\bf c}^T {\bf x} = {\bf x}^T diag({\bf c}) {\bf x}$ without affecting optimality. On the other hand, in a similar manner as we proceeded in (\ref{e4}) adding zero-valued terms, we employ an equivalent representation to accomplish the same result. Thus, the sum of weighted pair-wise products $c_{ij} x_{ij} x_{ik}$ with $r_j, r_k \in \mathcal{R}_{\alpha}$, can be expressed as ${\bf x}^T \big( {\bf I}_{M \times M} \otimes [{\bf 1}_{K \times K} - {\bf I}_{K \times K}] \big) diag({\bf c}) {\bf x} = 0$. Now, we are able to augment the cost function in (7a) and recast it as follows
\begin{equation} \label{e8}
	\small
	\hspace{0.1cm}
	\begin{array}{l}
	~ {\bf c}^T {\bf x} \\
	
	\vspace{-0.2cm} \\
	
	= {\bf x}^T diag({\bf c}) {\bf x} \\
	
	\vspace{-0.2cm} \\
	
	= {\bf x}^T  diag({\bf c}) {\bf x} + {\bf x}^T ({\bf I}_{M \times M} \otimes [{\bf 1} - {\bf I}]_{K \times K}) ~  diag({\bf c}) {\bf x}\\
	
	\vspace{-0.2cm} \\
	
	= {\bf x}^T ({\bf I}_{M \times M} \otimes {\bf I}_{K \times K} + {\bf I}_{M \times M} \otimes [{\bf 1} - {\bf I}]_{K \times K})  diag({\bf c}) {\bf x}\\

	\vspace{-0.2cm} \\
	
	= {\bf x}^T ({\bf I}_{M \times M} \otimes {\bf 1}_{K \times K})  diag({\bf c}) {\bf x}\\
	
	\end{array}
\end{equation}
\begin{mdframed}[] 
	\textbf{Property 1 (Product of two tensor products)}\\
	Let ${\bf X} \in \mathbb{R}^{m \times n}$, ${\bf Y} \in \mathbb{R}^{r \times s}$, ${\bf W} \in \mathbb{R}^{n \times p}$, and ${\bf Z} \in \mathbb{R}^{s \times t}$, then
	\begin{center}
		${\bf XY} \otimes {\bf WZ} = ({\bf X} \otimes {\bf W})({\bf Y} \otimes {\bf Z}) \in \mathbb{R}^{mr \times pt}$
	\end{center}
\end{mdframed}

Employing Property 1, (\ref{e8}) can be further simplified. Thus, the resultant cost function is shown in (\ref{e9})
\begin{equation} \label{e9}
	\hspace{0.1cm}
	\begin{array}{l}
		~ {\bf x}^T ({\bf I}_{M \times M} \otimes {\bf 1}_{K \times K}) diag({\bf c}) {\bf x}\\
		
		\vspace{-0.2cm} \\
		
		= {\bf x}^T ({\bf I}_{M \times M} {\bf I}_{M \times M} \otimes {\bf 1}_{K \times 1} {\bf 1}_{1 \times K}) diag({\bf c}) {\bf x}\\
		
		\vspace{-0.2cm} \\
		
		= \underbrace{{\bf x}^T ({\bf I}_{M \times M} \otimes {\bf 1}_{K \times 1})}_{\text{\bf y}^T} 
		\underbrace{({\bf I}_{M \times M} \otimes {\bf 1}_{1 \times K}) diag({\bf c}) {\bf x}}_{\text{\bf d}} \\
	
	\end{array}
\end{equation}

From (\ref{e9}), we obtain that ${\bf d} = ({\bf I}_{M \times M} \otimes {\bf 1}_{1 \times K}) diag({\bf c}) {\bf x}$ and ${\bf y} = ({\bf I}_{M \times M} \otimes {\bf 1}_{1 \times K}) {\bf x}$. Hence, we find that ${\bf x} = ({\bf I}_{M \times M} \otimes {\bf 1}_{1 \times K})^{\dagger}{\bf y}$. 
\begin{mdframed}[] 
	\textbf{Property 2 (Pseudo-inverse of a tensor product)}\\
	Let ${\bf X} \in \mathbb{R}^{m \times n}$ and ${\bf Y} \in \mathbb{R}^{r \times s}$, then
	\begin{center}
		$({\bf X \otimes Y})^{\dagger} = {\bf X}^{\dagger} \otimes {\bf Y}^{\dagger} \in \mathbb{R}^{ns \times mr}$
	\end{center}
\end{mdframed}
\begin{figure}[!b]
	\centering
	\begin{tikzpicture}
	
	\draw (1,0) rectangle (3.5, -0.6) node[pos=.5] {${\bf I}_{M \times M}\otimes {\bf 1}_{1 \times K}$};
	\draw (1,-1) rectangle (3.5, -1.5) node[pos=.5] {${\bf I}_{M \times M}\otimes {\bf 1}_{1 \times K}$};
	\draw (-0.8,-1.1) rectangle (-0.5, -1.4) node[pos=.5] {$\times$};
	\draw (-2.5,-1) rectangle (-1.3, -1.5) node[pos=.5] {$diag(\cdot)$};
	
	\draw [->] (-3, -0.3) -- (1, -0.3);
	\draw [->] (-0.5, -1.25) -- (1, -1.25);
	\draw [->] (-0.65, -0.3) -- (-0.65, -1.1);
	\draw [->] (-1.3, -1.25) -- (-0.8, -1.25);
	\draw [->] (-3, -1.25) -- (-2.5, -1.25);
	\draw [->] (3.5, -0.3) -- (4, -0.3);
	\draw [->] (3.5, -1.25) -- (4, -1.25);
	
	\node[text width = 0.2cm] at (-3.2,-0.3) {$\bf x$};
	\node[text width = 0.2cm] at (-3.2,-1.25) {$\bf c$};
	\node[text width = 0.2cm] at (4.2,-0.3) {$\bf y$};
	\node[text width = 0.2cm] at (4.2,-1.25) {$\bf d$};
	
	\end{tikzpicture}
	\caption{Transformation Process}
	\label{fig4}
\end{figure}
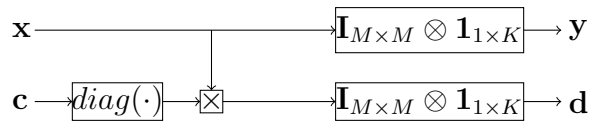
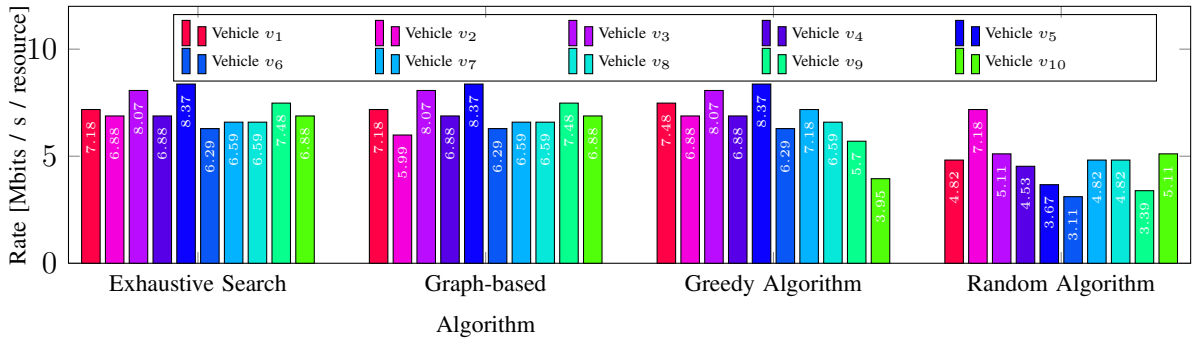
\begin{figure*}[!t]
	\centering
	\begin{tikzpicture}
	\begin{axis}[
	ybar,
	ymin = 0,
	ymax = 10,
	width = 16.5cm,
	height = 5cm,
	bar width = 7pt,
	tick align = inside,
	x label style={align=center, font=\footnotesize,},
	ylabel = {Rate [Mbits / s / resource]},
	y label style={at={(-0.025,0.5)}, font=\footnotesize,},
	nodes near coords,
	every node near coord/.append style={color = white, rotate = 90, anchor = east, font = \fontsize{2}{2}\selectfont},
	nodes near coords align = {vertical},
	symbolic x coords = {Exhaustive Search, Graph-based Algorithm, Greedy Algorithm, Random Algorithm},
	x tick label style = {text width = 3cm, align = center, font = \footnotesize,},
	xtick = data,
	enlarge y limits = {value = 0.2, upper},
	enlarge x limits = 0.15,
	legend columns=5,
	legend pos = north east,
	legend style={font=\fontsize{6}{5}\selectfont, text width=2.1cm,text height=0.02cm,text depth=.ex, fill = none, }]
	]
	
	\addplot[fill = color0] coordinates {(Exhaustive Search, 7.18) (Graph-based Algorithm, 7.18) (Greedy Algorithm, 7.48) (Random Algorithm, 4.82)}; \addlegendentry{Vehicle $v_{1}$}
	
	\addplot[fill = color1] coordinates {(Exhaustive Search, 6.88) (Graph-based Algorithm, 5.99) (Greedy Algorithm, 6.88) (Random Algorithm, 7.18)}; \addlegendentry{Vehicle $v_2$}
	
	\addplot[fill = color2] coordinates {(Exhaustive Search, 8.07) (Graph-based Algorithm, 8.07) (Greedy Algorithm, 8.07) (Random Algorithm, 5.11)}; \addlegendentry{Vehicle $v_3$}
	
	\addplot[fill = color3] coordinates {(Exhaustive Search, 6.88) (Graph-based Algorithm, 6.88) (Greedy Algorithm, 6.88) (Random Algorithm, 4.53)}; \addlegendentry{Vehicle $v_4$}
	
	\addplot[fill = color4] coordinates {(Exhaustive Search, 8.37) (Graph-based Algorithm, 8.37) (Greedy Algorithm, 8.37) (Random Algorithm, 3.67)}; \addlegendentry{Vehicle $v_5$}
	
	\addplot[fill = color5] coordinates {(Exhaustive Search, 6.29) (Graph-based Algorithm, 6.29) (Greedy Algorithm, 6.29) (Random Algorithm, 3.11)}; \addlegendentry{Vehicle $v_6$}
	
	\addplot[fill = color6] coordinates {(Exhaustive Search, 6.59) (Graph-based Algorithm, 6.59) (Greedy Algorithm, 7.18) (Random Algorithm, 4.82)}; \addlegendentry{Vehicle $v_7$}
	
	\addplot[fill = color7] coordinates {(Exhaustive Search, 6.59) (Graph-based Algorithm, 6.59) (Greedy Algorithm, 6.59) (Random Algorithm, 4.82)}; \addlegendentry{Vehicle $v_8$}
	
	\addplot[fill = color8] coordinates {(Exhaustive Search, 7.48) (Graph-based Algorithm, 7.48) (Greedy Algorithm, 5.70) (Random Algorithm, 3.39)}; \addlegendentry{Vehicle $v_9$}
	
	\addplot[fill = color9] coordinates {(Exhaustive Search, 6.88) (Graph-based Algorithm, 6.88) (Greedy Algorithm, 3.95) (Random Algorithm, 5.11)}; \addlegendentry{Vehicle $v_{10}$}
	
	\end{axis}
	\end{tikzpicture}
	\caption{One-shot simulation for different approaches}
	\label{fig5}
	\vspace{-0.2cm}
\end{figure*}

Employing Property 2, we obtain that ${\bf x} = {\bf I}_{M \times M}^{\dagger} \otimes {\bf 1}_{1 \times K}^{\dagger}~{\bf y}$. In the following, we use the previous relation in order to simplify the constraint (7b),
\begin{equation} \label{e10}
	\hspace{0.cm}
	\small
	\begin{array}{l}
	
		~\left( \left[
		\begin{array}{c}
			{\bf I}_{N \times N} \otimes {\bf 1}_{1 \times N}\\
			\hline
			{\bf 1}_{1 \times N} \otimes {\bf I}_{N \times N} 
		\end{array}
		\right] 
		\otimes {\bf 1}_{1 \times K} \right) {\left( {\bf I}_{M \times M} \otimes {\bf 1}_{1 \times K}^{\dagger} \right){\bf y}} = {\bf 1}
		
		\vspace{0.3cm} \\
		
		= \left( \left[
		\begin{array}{c}
			{\bf I}_{N \times N} \otimes {\bf 1}_{1 \times N}\\
			\hline
			{\bf 1}_{1 \times N} \otimes {\bf I}_{N \times N} 
		\end{array}
		\right] 
		{\bf I}_{M \times M} \right) \otimes \underbrace{{\left( {\bf 1}_{1 \times K} {\bf 1}_{1 \times K}^{\dagger} \right)}}_\text{1} {\bf y} = {\bf 1}
		
		\vspace{0.01cm} \\
		
		= \left[
			\begin{array}{c}
			{\bf I}_{N \times N} \otimes {\bf 1}_{1 \times N}\\
			\hline
			{\bf 1}_{1 \times N} \otimes {\bf I}_{N \times N} 
		\end{array}
		\right] 
		{\bf y} = {\bf 1}
	
	\end{array}
\end{equation}

Thus, the problem in (\ref{e7}) can be recast as (\ref{e11})
\begin{equation} \label{e11}
\hspace{-1.5cm}
\begin{array}{lclcl}
&& {\rm max} ~ {\bf d}^T {\bf y} \\
&& {\rm subject~to}~ 
\underbrace{
	{\left[
		\begin{array}{c}
		{\bf I}_{N \times N} \otimes {\bf 1}_{1 \times N}\\
		\hline
		{\bf 1}_{1 \times N} \otimes {\bf I}_{N \times N} 
		\end{array}
		\right]}}_\text{\bf {A}} {\bf y} = {\bf 1}.
\end{array}
\end{equation}

Since $\bf y$ is obtained from the product of binary variables $x_{ij}$ and a totally unimodular matrix $ ({\bf I}_{M \times M} \otimes {\bf 1}_{1 \times K})$, then ${\bf y} \in \mathbb{B}^M$. We can notice that (\ref{e11}) is identical to (\ref{e2}) but in terms of different variables. Fig.4 shows the transformation process from (\ref{e7}) to (\ref{e11}). Nevertheless, we can notice that $\bf d$ depends on $\bf x$ which is not desirable. In order to eliminate this dependency, we state without a proof---due to space limitations---that
\begin{equation} \label{e12}
	{\bf d} = \lim_{\beta \to \infty} \frac{1}{\beta} \overset{\substack{\circ}}{\log} \Big\{({\bf I}_{M \times M}\otimes {\bf 1}_{1 \times K}) \mathrm{e}^{\circ \beta {\bf c}} \Big\}
\end{equation}
where $\overset{\substack{\circ}}{\log} \{\cdot\}$ and $\mathrm{e}^{\circ \{ \cdot \} }$ are the element-wise natural logarithm and Hadamard exponential \cite{b8}, respectively.

\section{Simulations}
We consider a 10 MHz channel for conformity with ETSI ITS channelization \cite{b2}. The channel is divided into several resource chunks, each with an extent of 1 ms in time and 1.26 MHz in frequency. To wit, 1.26 MHz corresponds to 7 resource blocks (RBs), where one RB consists of 12 subcarriers spaced by 15 kHz \cite{b3}. The structure of each resource chunk is shown in Table I. Thus, in each subframe, there is a total of 7 resource chunks spanning 49 RBs. To the best of our understanding, a resource chunk with such proportions is sufficient to convey data (5RBs) and control information (2RBs). With appropriate modulation and coding schemes, CAM messages with a payload of 200 data bytes can be adequately supported in a resource chunk. The control information is necessary for compatibility with other vehicles that may be out of coverage (\textit{Mode 4}). Thus, these vehicles can be aware of the resources in use and self-allocate to themselves an idle resource.   
\begin{center}
	\begin{table}[h] \caption {Sidelink Resource Structure}
		\centering
		\begin{tabular}{l l}
			\toprule
			\multicolumn{1}{c}{\textbf{Description}} & \multicolumn{1}{c}{\textbf{Value}} \\
			\midrule
			Number of RBs & 7 \\
			Number of subcarriers per RB & 12 \\
			Number of data subcarriers & 60\\
			Number of control subcarriers & 24\\
			\bottomrule
		\end{tabular}
	\end{table}
	\vspace{-0.6cm}
\end{center}

In our model, we consider that clusters are totally independent from each other. This means that a subset of resources used in a certain cluster can be repurposed by vehicles in other clusters. The system parameters that we have employed in our simulations are detailed in Table II. Since we consider a message rate of 10 Hz and $N=100$ vehicles, the resource allocation task is carried out every 0.1 s. Signaling between vehicles and eNode via uplink/downlink resources should also be transmitted at least every 0.1 s.
\begin{center}
	\begin{table}[h] \caption {System Parameters}
		\centering
		\begin{tabular}{l l}
			\toprule
			\multicolumn{1}{c}{\textbf{Description}} & \multicolumn{1}{c}{\textbf{Value}} \\
			\midrule
			Number of vehicles per cluster & 10 - 100 \\
			Number of clusters & 1 - 7 \\
			Message rate (Hz) & 10\\
			Number of allottable subframes & 100\\
			Number of resources per subframe & 3, 7\\
			\bottomrule
		\end{tabular}
	\end{table}
	\vspace{-0.6cm}
\end{center}

In Fig. \ref{fig5} we show a one-shot simulation of the achievable rates comparing 4 different algorithms. For this particular simulation, we considered that the number of vehicles is $N = 10$ and the number of resources per subframe is $K = 3$. In the proposed \textit{graph-based algorithm} we have not enforced any mechanism to incentivize fairness. However, as can be observed, the approach can provide a fair resource assignment to all the vehicles. Furthermore, it attains the same results as \textit{exhaustive search}, which verifies its optimality. The \textit{greedy algorithm} can provide with good channel quality to some vehicles only but there are inevitably others with low quality conditions.

For the results in Fig. \ref{fig6}, we considered 4 communications clusters with $N = 100$ vehicles in each and $K = 7$. The results have been obtained in base of the average over 1000 simulations. We proved mathematically in Section IV that the proposed approach is optimal. Now, through simulations, we also show that our scheme can attain optimality as it achieves the same performance as \textit{exhaustive search} as can be observed in Fig. \ref{fig6}. Notice that \textit{greedy algorithm} performs as equally good as the proposed approach if we examine the highest-rate vehicle only. This is logical as the premise of the \textit{greedy algorithm} is assigning the best resources on first-come first-served basis. Considering the system average rate, our proposed approach has a small advantage. However, when considering the worst-rate vehicle, our proposal excels as it is capable of providing a higher level of fairness. In all cases, the \textit{random algorithm} is outperformed by the other approaches. 
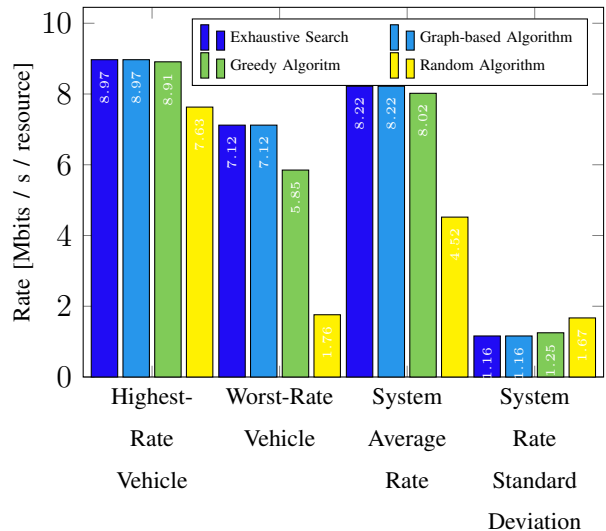
\begin{figure}[!t]
	\centering
	\begin{tikzpicture}
	\begin{axis}[
	ybar,
	ymin = 0,
	ymax = 9.5,
	width = 8.5cm,
	height = 6.5cm,
	bar width = 10pt,
	tick align = inside,
	x label style={align=center, font=\footnotesize,},
	ylabel = {Rate [Mbits / s / resource]},
	y label style={at={(-0.075,0.5)}, font=\footnotesize,},
	nodes near coords,
	every node near coord/.append style={color = white, rotate = 90, anchor = east, font = \fontsize{2}{2}\selectfont},
	nodes near coords align = {vertical},
	symbolic x coords = {Highest-Rate Vehicle, Worst-Rate Vehicle, System Average Rate, System Rate Standard Deviation},
	x tick label style = {text width = 1.6cm, align = center, font = \footnotesize,},
	xtick = data,
	enlarge y limits = {value = 0.1, upper},
	enlarge x limits = 0.18,
	legend columns=2,
	legend pos = north east,
	legend style={font=\fontsize{6}{5}\selectfont, text width=2.05cm,text height=0.02cm,text depth=.ex, fill = none, }
	]
	
	\addplot[fill = bblue] coordinates {(Highest-Rate Vehicle,  8.97) (Worst-Rate Vehicle, 7.12) (System Average Rate, 8.22) (System Rate Standard Deviation, 1.16)}; \addlegendentry{Exhaustive Search}
	\addplot[fill = sskyblue] coordinates {(Highest-Rate Vehicle, 8.97) (Worst-Rate Vehicle, 7.12) (System Average Rate, 8.22) (System Rate Standard Deviation, 1.16)}; \addlegendentry{Graph-based Algorithm}
	\addplot[fill = ggreen] coordinates {(Highest-Rate Vehicle, 8.91) (Worst-Rate Vehicle, 5.85) (System Average Rate, 8.02) (System Rate Standard Deviation, 1.25)}; \addlegendentry{Greedy Algoritm}
	\addplot[fill = yellow] coordinates {(Highest-Rate Vehicle, 7.63) (Worst-Rate Vehicle, 1.76) (System Average Rate, 4.52) (System Rate Standard Deviation, 1.67)}; \addlegendentry{Random Algorithm}

	\end{axis}
	\end{tikzpicture}
	\caption{Vehicles Data Rate}
	\label{fig6}
	\vspace{-0.2cm}
\end{figure}
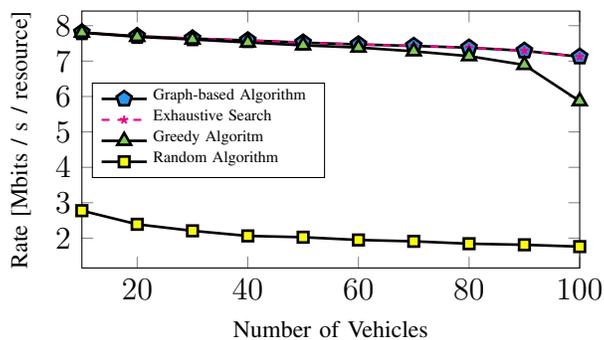
\begin{figure}[!t]
	\centering
	\begin{tikzpicture}
	\begin{axis}[
	xmin = 10,
	xmax = 100,
	width = 8.2cm,
	height = 5cm,
	xlabel={Number of Vehicles},
	x label style={align=center, font=\footnotesize,},
	ylabel = {Rate [Mbits / s / resource]},
	y label style={at={(-0.08,0.5)}, text width = 4cm, align=center, font=\footnotesize,},
	ytick = {1, 2, 3, 4, 5, 6, 7, 8},
	legend style={at={(0.02,0.35)},anchor=south west, font=\fontsize{6}{5}\selectfont, text width=2.1cm,text height=0.075cm,text depth=.ex, fill = none,},
	]
	\addplot[color=black, mark = pentagon*, mark options = {scale = 1.5, fill = sskyblue}, line width = 1pt] coordinates 
	{
		(10, 7.8084)
		(20, 7.7028)
		(30, 7.6374)
		(40, 7.5846)
		(50, 7.5216)
		(60, 7.4712)
		(70, 7.4280)
		(80, 7.3794)
		(90, 7.2912)
		(100, 7.1238)
	}; \addlegendentry{Graph-based Algorithm}

	\addplot[color=magenta, mark = star, mark options = {scale = 0.8}, line width = 0.8pt, style = dashed] coordinates 
	{
		(10, 7.8084)
		(20, 7.7028)
		(30, 7.6374)
		(40, 7.5846)
		(50, 7.5216)
		(60, 7.4712)
		(70, 7.4280)
		(80, 7.3794)
		(90, 7.2912)
		(100, 7.1238)
	}; \addlegendentry{Exhaustive Search}
	
	\addplot[color=black, mark = triangle*, mark options = {scale = 1.5, fill = ggreen}, line width = 1pt] coordinates 
	{
		(10, 7.8030)
		(20, 7.6848)
		(30, 7.6104)
		(40, 7.5264)
		(50, 7.44545)
		(60, 7.3830)
		(70, 7.2768)
		(80, 7.1400)
		(90, 6.8922)
		(100, 5.8686)
	}; \addlegendentry{Greedy Algoritm}
	
	\addplot[color=black, mark = square*, mark options = {fill = yyellow}, line width = 1pt] coordinates 
	{
		(10, 2.7762)
		(20, 2.3910)
		(30, 2.2074)
		(40, 2.0634)
		(50, 2.0262)
		(60, 1.9488)
		(70, 1.9104)
		(80, 1.8432)
		(90, 1.8126)
		(100, 1.7616)
	}; \addlegendentry{Random Algorithm}
	\end{axis}
	\end{tikzpicture}
	\caption{Worst-rate vehicle}
	\label{fig7}
	\vspace{-0.2cm}
\end{figure}

Fig. \ref{fig7} shows the achievable rate for the worst-rate vehicle. The proposed \textit{graph-based algorithm} attains the same performance as \textit{exhaustive search}. We observe that when the vehicle density per cluster is low, the \textit{greedy algorithm} attains near optimal solutions as there are far more resources than vehicles to serve. However, as the density increases, especially near the overload state, its performance drops. The \textit{random algorithm} performs worse than the other approaches.

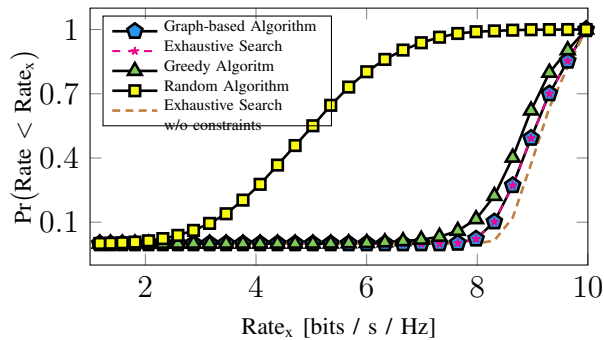
\begin{figure}[!]
	\centering
	\begin{tikzpicture}
	\begin{axis}[
	xmin = 1,
	xmax = 10,
	width = 8.2cm,
	height = 5cm,
	xlabel={Rate\textsubscript{x} [bits / s / Hz]},
	x label style={align=center, font=\footnotesize,},
	ylabel = {Pr\big(Rate~\textless~Rate\textsubscript{x}\big)},
	y label style={at={(-0.08,0.5)}, text width = 3.5cm, align=center, font=\footnotesize,},
	ytick = {0.1, 0.4, 0.7, 1.0},
	legend style={at={(0.025,0.54)},anchor=south west, font=\fontsize{6}{5}\selectfont, text width=2.05cm,text height=0.06cm,text depth=.ex, fill = none, align = left},
	]
	
	\addplot[color = black, mark = pentagon*, mark options = {scale = 1.5, fill = sskyblue, solid}, line width = 1pt] coordinates 
	{
		(1.1756, 0)
		(1.3701, 0)
		(1.5827, 0)
		(1.8122, 0)
		(2.0574, 0)
		(2.3165, 0)
		(2.5878, 0)
		(2.8698, 0)
		(3.1608, 0)
		(3.4594, 0)
		(3.7644, 0)
		(4.0746, 0)
		(4.3891, 0)
		(4.7070, 0)
		(5.0278, 0)
		(5.3509, 0)
		(5.6758, 0)
		(6.0022, 0)
		(6.3297, 0)
		(6.6582, 0)
		(6.9875, 0)
		(7.3173, 0)
		(7.6476, 0.0025)
		(7.9784, 0.0221)
		(8.3094, 0.1023)
		(8.6406, 0.2718)
		(8.9721, 0.4928)
		(9.3037, 0.7007)
		(9.6354, 0.8530)
		(9.9672, 1)
	}; \addlegendentry{Graph-based Algorithm}
	
	\addplot[color=magenta, mark = star, mark options = {scale = 0.8, fill = blue}, line width = 0.8pt, style = dashed] coordinates 
	{
		(1.1756, 0)
		(1.3701, 0)
		(1.5827, 0)
		(1.8122, 0)
		(2.0574, 0)
		(2.3165, 0)
		(2.5878, 0)
		(2.8698, 0)
		(3.1608, 0)
		(3.4594, 0)
		(3.7644, 0)
		(4.0746, 0)
		(4.3891, 0)
		(4.7070, 0)
		(5.0278, 0)
		(5.3509, 0)
		(5.6758, 0)
		(6.0022, 0)
		(6.3297, 0)
		(6.6582, 0)
		(6.9875, 0)
		(7.3173, 0)
		(7.6476, 0.0025)
		(7.9784, 0.0221)
		(8.3094, 0.1023)
		(8.6406, 0.2718)
		(8.9721, 0.4928)
		(9.3037, 0.7007)
		(9.6354, 0.8530)
		(9.9672, 1)
	}; \addlegendentry{Exhaustive Search}
	
	\addplot[color=black, mark = triangle*, mark options = {scale = 1.5, fill = ggreen}, line width = 1pt] coordinates 
	{
		(1.1756, 0)
		(1.3701, 0)
		(1.5827, 0)
		(1.8122, 0)
		(2.0574, 0)
		(2.3165, 0)
		(2.5878, 0)
		(2.8698, 0)
		(3.1608, 0)
		(3.4594, 0)
		(3.7644, 0)
		(4.0746, 0)
		(4.3891, 0)
		(4.7070, 0)
		(5.0278, 0.0001)
		(5.3509, 0.0005)
		(5.6758, 0.0019)
		(6.0022, 0.0031)
		(6.3297, 0.0064)
		(6.6582, 0.0115)
		(6.9875, 0.0187)
		(7.3173, 0.0322)
		(7.6476, 0.0597)
		(7.9784, 0.1143)
		(8.3094, 0.2230)
		(8.6406, 0.4017)
		(8.9721, 0.6218)
		(9.3037, 0.7983)
		(9.6354, 0.9032)
		(9.9672, 1)
	}; \addlegendentry{Greedy Algoritm}
	
	\addplot[color=black, mark = square*, mark options = {fill = yyellow, solid}, line width = 1pt] coordinates 
	{
		(1.1756, 0.0010)
		(1.3701, 0.0021)
		(1.5827, 0.0041)
		(1.8122, 0.0072)
		(2.0574, 0.0134)
		(2.3165, 0.0235)
		(2.5878, 0.0386)
		(2.8698, 0.0619)
		(3.1608, 0.0946)
		(3.4594, 0.1395)
		(3.7644, 0.2033)
		(4.0746, 0.2771)
		(4.3891, 0.3669)
		(4.7070, 0.4581)
		(5.0278, 0.5505)
		(5.3509, 0.6461)
		(5.6758, 0.7312)
		(6.0022, 0.8028)
		(6.3297, 0.8604)
		(6.6582, 0.9041)
		(6.9875, 0.9383)
		(7.3173, 0.9641)
		(7.6476, 0.9815)
		(7.9784, 0.9897)
		(8.3094, 0.9936)
		(8.6406, 0.9970)
		(8.9721, 0.9991)
		(9.3037, 0.9996)
		(9.6354, 0.9999)
		(9.9672, 1)
	}; \addlegendentry{Random Algorithm}
	
	\addplot[color=brown, mark options = {fill = yyellow}, line width = 1pt, style = densely dashed] coordinates 
	{
		(1.1756, 0)
		(1.3701, 0)
		(1.5827, 0)
		(1.8122, 0)
		(2.0574, 0)
		(2.3165, 0)
		(2.5878, 0)
		(2.8698, 0)
		(3.1608, 0)
		(3.4594, 0)
		(3.7644, 0)
		(4.0746, 0)
		(4.3891, 0)
		(4.7070, 0)
		(5.0278, 0)
		(5.3509, 0)
		(5.6758, 0)
		(6.0022, 0)
		(6.3297, 0)
		(6.6582, 0)
		(6.9875, 0)
		(7.3173, 0)
		(7.6476, 0)
		(7.9784, 0.0001)
		(8.3094, 0.0146)
		(8.6406, 0.1208)
		(8.9721, 0.3712)
		(9.3037, 0.6350)
		(9.6354, 0.8252)
		(9.9672, 1)
	}; \addlegendentry{Exhaustive Search \\ w/o constraints}
	\end{axis}
	\end{tikzpicture}
	\caption{Cumulative Distribution Function}
	\label{fig8}
	\vspace{-0.2cm}
\end{figure}

Fig. \ref{fig8} shows the cumulative distribution function (CDF) of the achievable rates. We observe that the proposed approach outperforms the other two approaches. For the sake of comparison, we have included the results of the unconstrained system, which does not takes into account conflict avoidance constraints. This is of course not desirable but it serves as a comparison bound.

The complexity of \textit{exhaustive search} is $\mathcal{O}({\vert \mathcal{R} \vert! / (\vert \mathcal{R} \vert - \vert \mathcal{V} \vert})!)$ whereas the complexity of \textit{graph-based algorithm} is $\mathcal{O}(\max \{ \vert \mathcal{V} \vert, \vert \mathcal{R} \vert / K \}^3)$. The complexity of the \textit{greedy algorithm} and \textit{random algorithm} are $\mathcal{O}({\vert \mathcal{V} \vert \vert \mathcal{R} \vert})$ and $\mathcal{O}({\vert \mathcal{V} \vert})$, respectively.

\section{Conclusion}
We have presented a novel resource allocation algorithm for V2V communications considering conflict constraints. We were able to transform the original problem into a simplified form by means of two approaches. In our future work, we will consider $(i)$ power control and $(ii)$ the assumption that a subset of vehicles may belong to more than one cluster simultaneously.

\end{document}